\documentclass[aps,pra,superscriptaddress,notitlepage,twocolumn]{revtex4-1}
\usepackage[utf8]{inputenc}
\usepackage{braket}
\usepackage{amsmath,amsfonts}
\usepackage{subcaption}
\usepackage{graphicx}
\usepackage{bm}
\usepackage{amsthm}
\usepackage[colorlinks]{hyperref}
\usepackage{xcolor}
\usepackage{comment}
\usepackage{mathtools}
\usepackage{multirow}
\usepackage{booktabs}

\newcommand{\ba}{\begin{eqnarray}}
\newcommand{\ea}{\end{eqnarray}}
\newcommand{\ban}{\begin{eqnarray*}}
\newcommand{\ean}{\end{eqnarray*}}
\usepackage{subcaption}

\newcommand{\valerio}[1]{{\color{blue} #1}}

\begin{document}

\title{Information causality beyond the random access code model}
\author{Baichu Yu}
\affiliation{Centre for Quantum Technologies, National University of Singapore, 3 Science Drive 2, Singapore 117543}
\affiliation{Shenzhen Institute for Quantum Science and Engineering,
Southern University of Science and Technology, Nanshan District, Shenzhen, 518055, China}
\affiliation{Quantum Science Center of Guangdong-Hong Kong-Macao Greater Bay Area, Shenzhen 518045, China}
\author{Valerio Scarani}
\affiliation{Centre for Quantum Technologies, National University of Singapore, 3 Science Drive 2, Singapore 117543}
\affiliation{Department of Physics, National University of Singapore, 2 Science Drive 3, Singapore 117542}

\begin{abstract}

Information causality (IC) was one of the first principles that have been invoked to bound the set of quantum correlations. For some families of correlations, this principle recovers exactly the boundary of the quantum set; for others, there is still a gap. We close some of these gaps using a new quantifier for IC, based on the notion of ``redundant information''. This progress was made possible by the recognition that the principle of IC can be captured without referring to the success criterion of random access codes. We give strong numerical evidence that the new definition is still obeyed by quantum correlations in the same scenario.

\end{abstract}

\maketitle

\textit{Introduction-} Quantum theory differs from classical theory by the fact that the space of states is a vector space, rather than a set. This change has been necessary to accommodate the fact that only a fraction of the possible physical properties can be well-defined in any given state. But why a vector space, instead of something else? The basic answer is pragmatic: it has worked amazingly well. But of course, it would be desirable to know which principles underlie this choice. In the last decades, it was understood that quantum entanglement plays a crucial role in addressing this question. All the recent representation theorems use an axiom that has to do with composite systems (see \cite{paviabook} for an overview).

The most direct signature of entanglement are the \textit{quantum correlations} obtained by measuring the sub-systems separately. It is well known that some of these correlations cannot be reproduced within classical theory without communication, because they violate Bell's inequalities \cite{scaranibook}. Instead of recovering the whole of quantum theory, a series of works have tried to find principles that single out the set $\mathcal{Q}$ of quantum correlations (save the phenomena, rather than the whole formalism). Since measurements on shared entanglement cannot be used to send a signal, Popescu and Rohrlich asked whether this \textit{no-signaling principle} singles out quantum correlations exactly: they quickly found that it does not \cite{PR94}. The question then became: can one find other principles, to add on top of no-signaling, so as to identify the set $\mathcal{Q}$?

Several such principles were proposed \cite{brassard06,pawlowski2009information,nw09,fritzLO}. Most of them have later been proved to be satisfied in the set of ``almost-quantum'' correlations \cite{almostq}, which is strictly larger than $\mathcal{Q}$. Thus, those principles are satisfied by all quantum correlations, but they don't identify $\mathcal{Q}$. For the \textit{principle of information causality (IC)} \cite{pawlowski2009information}, our knowledge is less definite. For sure, all quantum correlations satisfy it; but its relations with the set of almost-quantum correlations is not known, and it is then still possible that IC identifies $\mathcal{Q}$ for bipartite Bell correlations (some modifications will be needed for multipartite ones \cite{gallego2011,Yang_2012}). 
This possibility is further developed through recent studies, which improve the methods in obtaining IC boundaries for general bipartite quantum correlations \cite{jain2024information,gachechiladze2022quantum}, justify quantum composition rules \cite{patra2023principle} and derive communication complexity bounds \cite{miklin2026communicationcomplexityboundsinformation} using IC.

IC was defined in terms of a task: a classical random access code (RAC), augmented by sharing the no-signaling resource under study between the two players. In this paper, we propose to redefine IC in a way that captures the same underlying notion as the original, but without reference to the specific task of RAC. The new criterion is violated by a larger set of non-quantum correlations. To substantiate this claim, we show the first tightening of the IC boundary in the simplest Bell scenario (two parties, two inputs and two outputs, 2-2-2) since the original study \cite{allcock2009recovering}.
We give strong numerical evidence that the new IC definition is obeyed by all 2-2-2 quantum correlations under the same protocol. Numerical evidence also supports that our IC is a quasi-convex function on the set of correlations, which is a property with clear physical meaning.   



\textit{Reconsidering information causality.--} We use upper case letters $A,B$ to denote random variables, lower case letters $a,b$ to denote specific values. For bits, thus, one should understand $A=\{0,1\}$, and then $a=0$ or $a=1$. The probability distribution of the variable $X$ is denoted $p_{X}$, with $p_{X}(x)$ the probability of its event $x$ (we may sometimes omit the subscript for simplicity).

To define IC, one considers the following two-player game. At every round, Alice's input is a string of $N$ bits $A=A_1\times ...\times A_N$, its value $a=(a_1,...,a_N)$ drawn at random with uniform distribution. She can send information to Bob on a channel with capacity $k$ per round. In addition, she shares a no-signaling resource with Bob. If $k<N$, Bob obviously can retrieve at most $k$ bits of Alice' input, irrespective of what the no-signaling resource is. But with a clever use of some no-signaling resources, something unexpected may happen: Bob could choose \textit{which} $k$ bits to retrieve. In words, it looks like the information about all $N$ bits was ``potentially'' present at Bob's location, although eventually he can read out only $k$ of them. \textit{The principle of IC states that this should not happen; and more quantitatively: even the ``potential information'' available at Bob's location should not exceed the capacity $k$ of the channel linking Alice and Bob}. In this paper we discuss how this potential information should be quantified.

In the original paper \cite{pawlowski2009information} and all subsequent works \cite{allcock2009recovering,Barnum_2010,cavalcanti,alsafi,Yang_2012,miklin2021information}, IC was captured by a random access code (RAC) criterion: Bob receives an input $t\in\{1,...,N\}$ that tells him which of Alice's bits he is supposed to retrieve in any given round. Over many rounds, then, the potential information is estimated by
\begin{equation}\label{ICoriginal}
IC_{\textrm{RAC}}=\sum_{i=1}^{N}I(A_{i};B|t=i) \,\equiv\,\sum_{i=1}^{N}I(A_{i};B_i)\,.
\end{equation} With this criterion, IC is satisfied if 
\begin{equation}\label{ICRAC}
IC_{\textrm{RAC}}\leq k.
\end{equation}
The intuition behind such RAC type of IC characterization is: we would find it surprising if, in every round, Bob could produce the correct value $b_i=a_i$ for the requested $i$, thus achieving $IC_{\textrm{RAC}}=N$.

But the definition of potential information does not necessarily require a game with a specific winning criterion for each round. The input $A$ of Alice can be treated as a single symbol (which doesn't even need to have the dimension of a string of bits), and the number $M$ of Bob's inputs can be an independent number (even if Alice's input were $N$ bits, we could have $M\neq N$). After many rounds, Alice and Bob can estimate the $M$ probability distributions $p_{AB}(a,b|t=i)\equiv p_{AB_i}(a,b_i)$. A bound on potential information can be obtained directly from those. Specifically, the quantifier of potential information that we consider would read
\begin{equation}\label{ICnew}
IC_{\textrm{red}}(M)=\sum_{i=1}^{M}I(A;B_i)-I_r(A;B_1,...,B_M)\,.
\end{equation}
The first term in \eqref{ICnew} is the sum of the mutual information of $A$ with each of the $B_i$. It can easily reach $Mk$ if $B_1$ carries all the information $k$, and all the other $B_i$ are set equal to $B_1$ and thus carry the same piece of information. This observation is the basis for understanding the role of the second term: one needs to \textit{remove redundant information}, i.e.~information that is present in several $B_i$. The characterization of redundant information is still debated. The lack of a general expression for $I_r(A;B_1,...,B_M)$ is the current limit for our study of IC. Fortunately, the interest of the approach can already be proved in the simplest case $M=2$, for which an expression for redundant information has been given.

\textit{IC with Redundant Information.--} We are going to study
\begin{align}\label{modifiedIC}
IC_{\textrm{red}}= \sum_{i=1}^{2}I(A;B_{i})-I_{r}(A;B_{1},B_{2})
\end{align}
with the measure of redundant information proposed by Harder, Salge and Polani~\cite{harder2013bivariate}. To compute $I_{r}(A;B_{1},B_{2})$, the starting point are the two joint probability distributions $p_{AB_i}$ for $i=1,2$. From $p_{AB_i}$, for each value of $B_i$ one constructs the probability distribution $p_{A|b_i}$ on $A$. These are $|B_i|$ points in the probability simplex of $A$, and we denote by $C_i$ their convex hull. For $|B|=2$, as is the case for us, 
\ba C_i=\big\{\lambda p_{A|b_i=0}+(1-\lambda)p_{A|b_i=1}\,|\,\lambda\in [0,1]\big\}\,.\label{convex}\ea
Now, for every value $b_1$ of $B_1$, one defines $p_{A[b_1\searrow B_2]}$ as the element of $C_2$ that is ``closest'' to $p_{A|b_1}$ in the following sense: 
\begin{equation}\label{argmin}
p_{A[b_1\searrow B_2]}:=\textrm{argmin}_{r\in C_2} D_{KL}(p_{A|b_1}||r),
\end{equation}
where $D_{KL}$ is the Kullback-Leibler divergence. Having solved this optimisation for all $b_1$, one can compute the ``projected information'' \footnote{The projected information can be written in a more telling way as
\[
I_{A}^{\pi}(B_1\searrow{B_2})=I(A;B_1)-D_{KL}(p_{AB_1}||\tilde{p}_{AB_1})
\]
where $\tilde{p}_{AB_1}(a,b_1)=p_{A[b_1\searrow B_2]}(a)p_{B_1}(b_1)$ could be understood as the best guess for the distribution of $(A,B_1)$ given $(A,B_2)$. In order to go from \eqref{projection} to this expression, one multiplies both the numerator and the denominator inside the logarithm by $p_{B_1}(b_1)p_{AB_1}(a,b_1)$ and rearranges the terms. (Francesco Buscemi, private communication).
}
\begin{equation}\label{projection}
I_{A}^{\pi}(B_1\searrow{B_2})=\sum_{a,b_1}p_{AB_1}(a,b_1)\log\left(\frac{p_{A[b_1\searrow B_2]}(a)}{p_A(a)}\right)\,.
\end{equation}
After repeating the recipe with $B_1$ and $B_2$ exchanged, redundant information is finally computed as
\begin{equation}\label{redundant}
I_{r}(A;B_1,B_2):=\min\{I_{A}^{\pi}(B_1\searrow B_2),I_{A}^{\pi}(B_2\searrow B_1)\}\,.
\end{equation}

\textit{Evidence of improvement.--} \label{secresults} 
IC expression of the form \eqref{modifiedIC} takes into account the global patterns of $A$ described by each $B_{i}$ instead of bit-to-bit correspondence, which provides it more power in some scenarios.

We now prove that our approach to IC constitutes a real improvement over the original one, as it rules out more non-quantum correlations. Specifically, we are going to report the first improvement on IC for the 2-2-2 scenario since the original study by Allcock and coworkers \cite{allcock2009recovering}. 


We denote $x,y\in\{0,1\}$ the inputs to the boxes, $\mathfrak{a},\mathfrak{b}\in\{0,1\}$ the outputs. The set of no-signaling correlation is a polytope with 24 vertices: 8 \textit{extremal non-local boxes}, which can be parametrized by $\mu,\nu,\sigma\in\{0,1\}$ as
\begin{equation}
P_{NL}^{\mu\nu\sigma}(\mathfrak{a}\mathfrak{b}|xy)=
\left \{
\begin{array}{ll}
\frac{1}{2},\ &\textrm{if}\ \mathfrak{a}\oplus \mathfrak{b}=xy\oplus\mu x\oplus \nu y\oplus \sigma \\
0,&\textrm{otherwise}\,;
\end{array}
\right.
\end{equation}
and 16 \textit{local deterministic boxes}, which can be parametrized by $\mu,\nu,\sigma,\tau\in\{0,1\}$ as
\begin{equation}
P_{L}^{\mu\nu\sigma\tau}(\mathfrak{a}\mathfrak{b}|xy)=
\left \{
\begin{array}{ll}
1,\ &\textrm{if}\ \mathfrak{a}=\mu x\oplus \nu,\ \mathfrak{b}=\sigma y\oplus \tau\\
0,&\textrm{otherwise}.
\end{array}
\right.
\end{equation}
The box $P_{NL}^{000}$ is the canonical form of the PR-box, maximizing the value of $CHSH=C_{00}+C_{01}+C_{10}-C_{11}$ where $C_{xy}=P(\mathfrak{a}=\mathfrak{b}|xy)-P(\mathfrak{a}\neq\mathfrak{b}|xy)$. All the other  $P_{NL}^{\mu\nu\sigma}$ are obtained from it by relabelling some of the inputs and/or some of the outputs, and maximize the corresponding CHSH-type expression.

No method is known to decide whether a no-signaling resource violates IC based on the description of the box alone: one needs to invent an explicit protocol that uses that resource (the fact that this protocol may not be optimal is the main reasons why the exact boundaries of the violation of IC are not known). Here, we follow a recently proposed compact protocol \cite{miklin2021information}. Although it led to some improvements for other scenarios, for the 2-2-2 scenario this protocol reproduced the results of \cite{allcock2009recovering}. Thus, the improvement we are going to report is really due to our new definition of IC. 

The protocol is the following. Alice inputs $x=a_{1}\oplus a_{2}$ into the box; upon receiving the output $\mathfrak{a}$, she computes the bit $m=\mathfrak{a}\oplus a_{1}$. She sends this bit to Bob on a noisy channel: specifically, a symmetric binary channel that flips the bit with probability $1-p_c$. The capacity of this channel is
\begin{equation}\label{capacity}
k=1+p_{c}\log_{2}p_{c}+(1-p_c)\log_2(1-p_c)\,.
\end{equation} 
In this paper, we set $p_{c}=0.5001$ since the bounds for IC get tighter in the limit $p_c\rightarrow \frac{1}{2}$ \cite{miklin2021information}.

On his side, when Bob wants to estimate $B_i$, he inputs $y=i-1$ in the box. Upon receiving the output $\mathfrak{b}$, he produces $b_i=\mathfrak{b}\oplus m'$ where $m'$ is the output of the noisy channel from Alice.

\textit{Case studies-} As concrete case studies, we look at the same three families of boxes studied in \cite{allcock2009recovering}. These families are defined by the convex combination
\begin{equation}\label{box}
PR_{\alpha,\beta}=\alpha P_{NL}^{000}+\beta R_{NS}+(1-\alpha-\beta) \mathbb{I},
\end{equation}
where $\alpha\in[0,1]$, $\beta\in[0,1-\alpha]$,  $R_{NS}$ is an extremal point of the no-signaling polytope, and $\mathbb{I}$ indicates the white noise $P(\mathfrak{a}\mathfrak{b}|xy)=\frac{1}{4}$. The case studies will involve the same three choices of $R_{NS}$ as \cite{allcock2009recovering}, namely $P_{NL}^{010}$, $P_{NL}^{110}$ and $P_{L}^{0000}$. By the symmetry of the problem, this choice covers actually all cases: $P_{NL}^{010}$ is equivalently to $P_{NL}^{011}$,  $P_{NL}^{100}$, and $P_{NL}^{101}$; $P_{NL}^{110}$ is equivalent to $P_{NL}^{111}$; and all the local deterministic points on the facet $CHSH=2$ are equivalent. Finally, since $P_{NL}^{001}=2\mathbb{I}-P_{NL}^{000}$ is the PR-box opposite to the canonical one, that mixing is already taken into account in \eqref{box}.

The explicit expressions needed to compute the $p_{AB_i}$ for each case are given in Table \ref{tab:probs}. The numerical calculations require setting $p_c$ close to $\frac{1}{2}$ because the tightest bounds are found in that limit \cite{miklin2021information}; and identifying the minimum in \eqref{argmin}. These are reliably dealt with by sampling evenly and by varying some precision parameters, details are given in footnote \footnote{For each curve, we choose 50 values of $\beta$ evenly distributed from $[0,1]$, and then 3000 values of $\alpha$ evenly distributed in $[0,1-\beta]$. We identify the largest $\alpha$ such that the IC do not exceed the channel capacity \eqref{capacity} with $p_{c}=0.5001$; we have run checks on some points with $p_c=0.50001$ and observed no visible difference. The optimisation \eqref{argmin} is also done by even sampling. Indeed, for each value of $b_1$, we have to find the value of $\lambda$ such that $r_A(\lambda)=\lambda p_{A|b_2=0}+(1-\lambda)p_{A|b_2=1}$ minimizes $D_{KL}(p_{A|b_1}||r_A(\lambda))$; then we have to redo this with the roles of $B_1$ and $B_2$ reversed. In each case, we sample $\lambda$ in the interval $[0,1]$ by steps of $\frac{1}{n}$ and identify the minimum. By inspection, we found that $n=1000$ was well sufficient (although we run checks for higher values of $n$, up to $n=20000$).}. As a check, we also redid the curves for the original IC criterion and recovered the plots of \cite{allcock2009recovering} as expected.

\begin{figure}[h]
    \centering
  \includegraphics[width=0.8\linewidth]{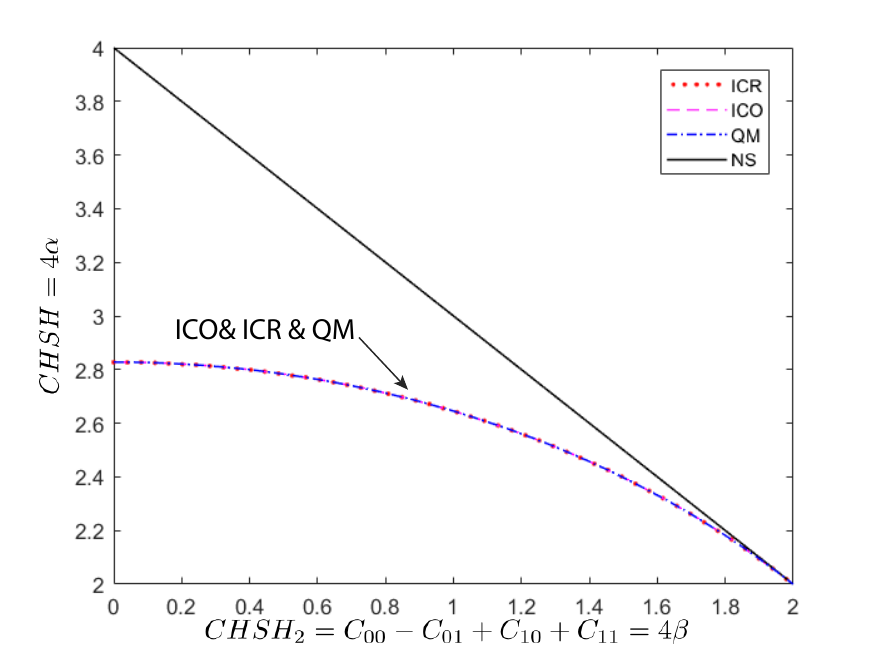}\\
  \includegraphics[width=0.8\linewidth]{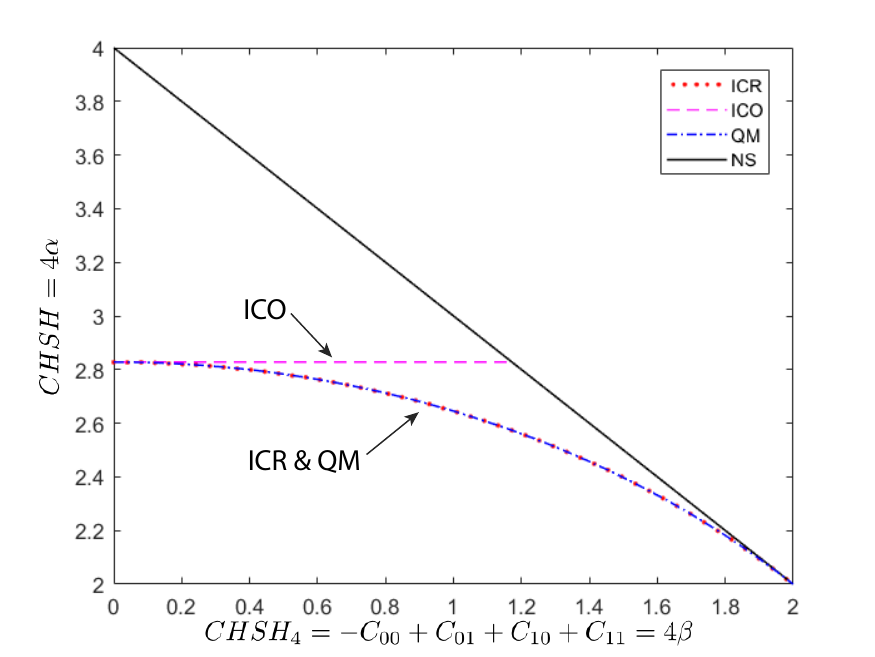}\\
  \includegraphics[width=0.8\linewidth]{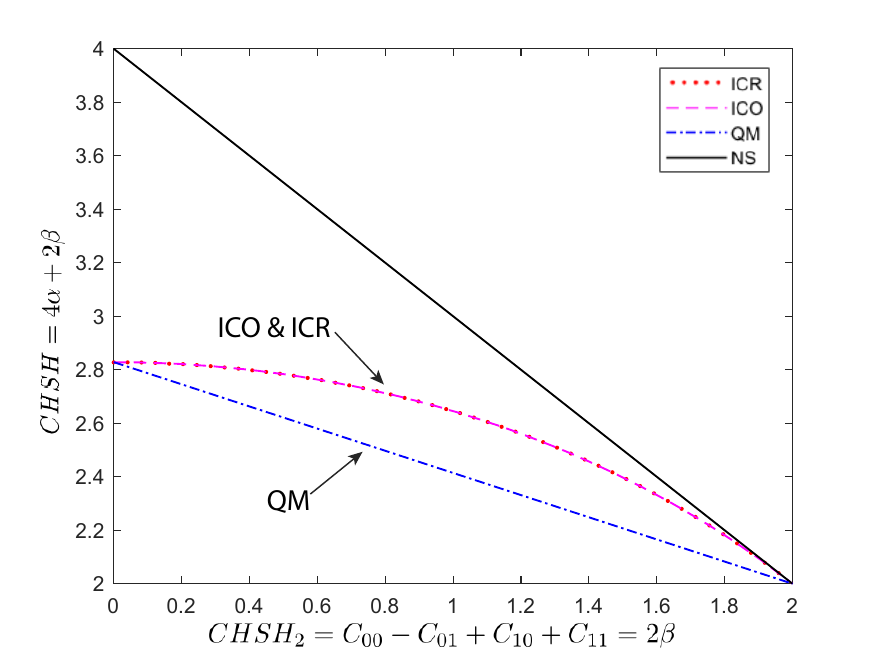}
  \caption{Slices \eqref{box} of the non-signalling polytope studied in this work: from top to bottom, $R_{NS}=P_{NL}^{010}$, $R_{NS}=P_{NL}^{110}$, and $R_{NS}=P_{L}^{0000}$. In all figures, the top left corner is the PR box $P_{NL}^{000}$, the bottom line is the facet $CHSH=2$. ICO represents IC for the original definition \eqref{ICoriginal}; these are the curves found in \cite{allcock2009recovering}. ICR represents IC for our definition based on redundant information \eqref{modifiedIC}. The quantum boundary is the Tsirelson-Landau-Masanes bound (see \cite{scaranibook}) for the first two figures, and a straight line for the third \cite{goh2018geometry}.}
  \label{fig:results}
  \end{figure}

The results are shown graphically in Fig.~\ref{fig:results}. In summary:
\begin{itemize}
    \item For $R_{NS}=P_{NL}^{010}$ (or $P_{NL}^{011}$,  $P_{NL}^{100}$, $P_{NL}^{101}$), both the original definition of IC and our new one recover the boundary of the quantum set (within numerical precision).
    \item $R_{NS}=P_{NL}^{110}$ (or $P_{NL}^{111}$), the original definition of IC stayed very far from the quantum boundary: in fact, it could just detect a violation of IC for those boxes that violate the Tsirelson bound. By contrast, our new definition recovers exactly the quantum set, within numerical precision.
    \item Finally, for $R_{NS}=P_{L}^{0000}$ (or any other local deterministic point on the facet $CHSH=2$), our definition and the original one give the same boundary for IC, but there remains a gap with the quantum boundary.
\end{itemize}

The fact that the original definition recovered the quantum boundary for some $R_{NS}=P_{NL}^{\mu\nu\sigma}$ but not others was an artefact of the use of RAC. This can be intuited by looking at the behavior of the extremal points for $p_c=1$. For the canonical PR-box $P_{NL}^{000}$, the protocol yields $A_1=B_1$ and $A_2=B_2$; for $R_{NS}=P_{NL}^{110}$, the same protocol yields $A_1=B_2\oplus 1$ and $A_2=B_1$. The amount of potential information is the same for both points (each $B_i$ has full information on one of the $A_j$), and our definition captures this. Imposing the RAC winning condition $A_i=B_i$ breaks the symmetry.

When $R_{NS}=P_{L}^{0000}$, the quantum boundary is provably a straight line \cite{goh2018geometry}. The boundary of the set of ``almost-quantum'' correlations $\mathcal{Q}_{1+AB}$ is indistinguishable from it at the scale of the figure \footnote{The boundary for $\mathcal{Q}_{1+AB}$ was computed by Koon Tong Goh (private communication) with the SeDuMi SDP solver. The difference between those values and the straight line is found at the third significant digit. Since the precision of the solver was set at with a precision $\sim 10^{-10}$, the difference is believed to be real.}. A significant gap remains between those sets and the violation of IC, even with the new definition that here does not improve on the original one. This gap may be real: definitely, this is a slice where one could focus the efforts to prove that IC does not coincide with $\mathcal{Q}$. Alternatively, we may not have captured ``potential information'' at its tightest yet. For instance, Ref.~\cite{ince2017measuring} argues that redundant information may have to depend also on the joint distribution of Bob's outputs $p_{B_1B_2}$ and not just on the marginals with Alice $p_{AB_i}$.

\begin{table}[h]
    \centering
\begin{tabular}{|c|cc|cc|cc|}
\hline
     $R_{NS}$& $P^{010}_{NL}$ && $P^{110}_{NL}$ && $P^{0000}_{L}$ & \\
     & $B_1$ & $B_2$ & $B_1$ & $B_2$ & $B_1$ & $B_2$ \\ \hline\hline
     $a=00$ &$k_+$&$k_-$&$k_+$&$k_-$&$k_+$&$k_+$\\ \hline
     $a=01$ &$k_+$&$1-k_-$&$k_-$&$1-k_+$&$k_+$&$1-k_-$\\ \hline
     $a=10$ &$1-k_+$&$k_-$&$1-k_-$&$k_+$&$1-k_+$&$k_-$\\ \hline
     $a=11$ &$1-k_+$&$1-k_-$&$1-k_+$&$1-k_-$&$1-k_+$&$1-k_+$\\ \hline
\end{tabular}
\caption{Values of $p(b_i=0|a)$ for the three slices \eqref{box} under study. We have denoted $k_\pm=\frac{1}{2}+(p_c-\frac{1}{2})(\alpha\pm\beta)$. Since the protocol has $p(a)=\frac{1}{4}$, and since in those slices $p(b_i)=\frac{1}{2}$, it holds $p(a|b_i)=p(b_i|a)/2$. The latter numbers define the $p_{A|b_i}$ used for the calculation of redundant information.}
\label{tab:probs}
\end{table}

\textit{Properties of IC.--}
Now we study two important properties about our new IC definition.
Firstly, any justified IC principle should always be satisfied by quantum correlations. That is, condition 
\begin{equation}\label{ineqICred}
    IC_{\textrm{red}}\leq k
\end{equation}
should be satisfied by all quantum correlations.
Secondly, we also argue that when the distribution of the variable $A$ is fixed,
an IC formalism should be a quasi-convex function on the set of correlations.  
A function \( f: C \to \mathbb{R} \) on a convex set $C$ is called quasi-convex if for all \( x, y \in C \) and all \( \lambda \in [0,1] \),  

\begin{equation}
 f\bigl(\lambda x + (1-\lambda) y \bigr) \le \max \{ f(x), f(y) \}.   \end{equation}  
Quasi-convexity guarantees that mixing resources does not improve the performance. 
For example, the original expression, $IC_{\text{RAC}}$, is convex (thus quasi-convex) over the distributions $p(A_i, B_i)$ for a fixed $p(A)$. 
Consequently, it remains convex over the set of correlations provided the protocol mapping those correlations to $p(A_i, B_i)$ is linear.

We provide numerical evidence of these two properties for the 2-2-2 scenario in two different approaches (we comment in the conclusion on the difficulties of finding an analytical proof). 
In the first approach,
We generated the set of 2-2-2 quantum correlations by characterizing the extremal points using a recent result in Ref.~\cite{barizien2025quantum}, and made random mixtures of these extremal correlations. 
Then we checked the $IC_{red}$ values that each correlation point produces under the same protocol we just adopted to obtain correlation boundaries.

\begin{figure}[htbp]
    \centering
    \begin{subfigure}[b]{0.9\linewidth}
        \centering
        \includegraphics[width=\linewidth]{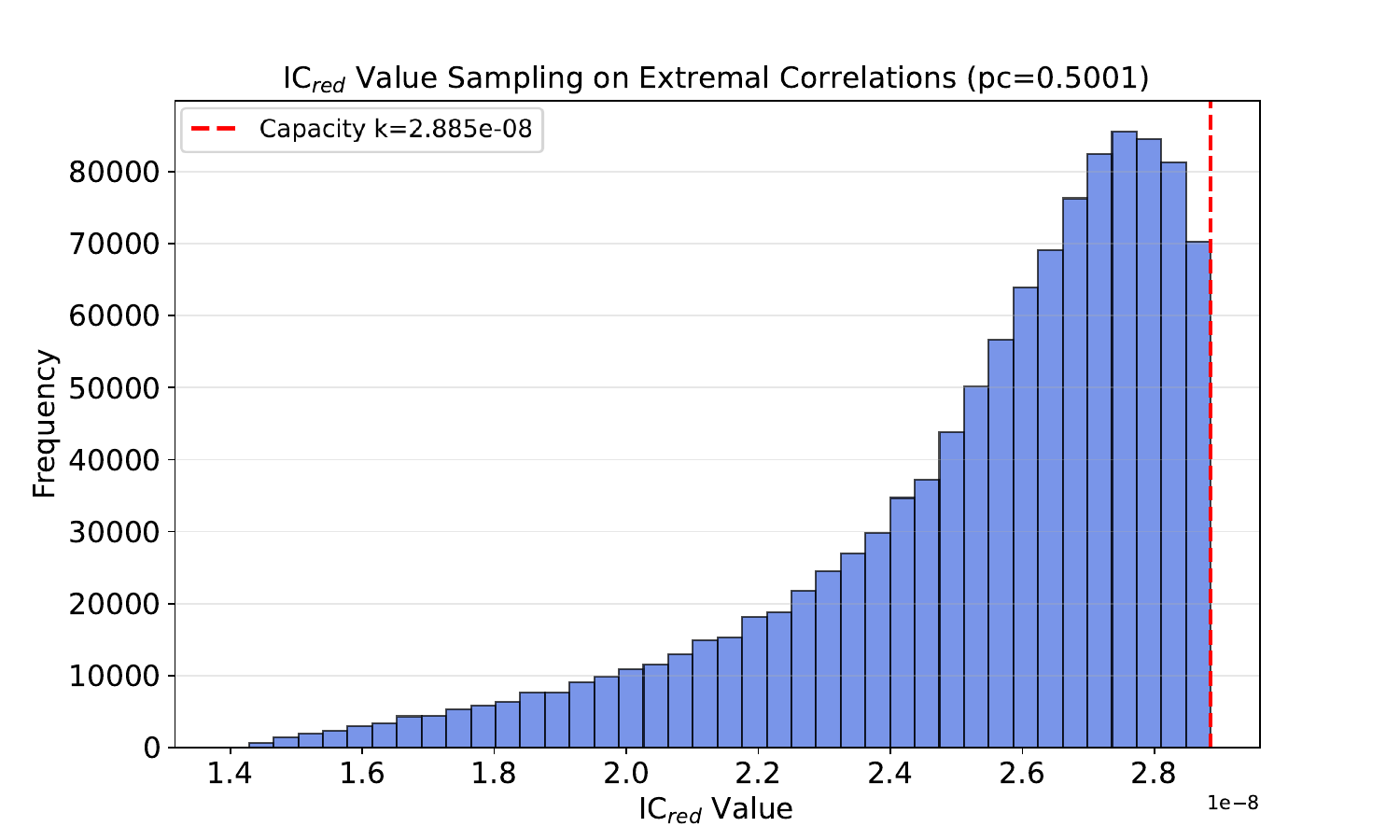}
        \caption{Distribution of $IC_{\mathrm{red}}$ values.}
        \label{sub:ic_red}
    \end{subfigure}
    \hfill 
    \begin{subfigure}[b]{0.9\linewidth}
        \centering
        \includegraphics[width=\linewidth]{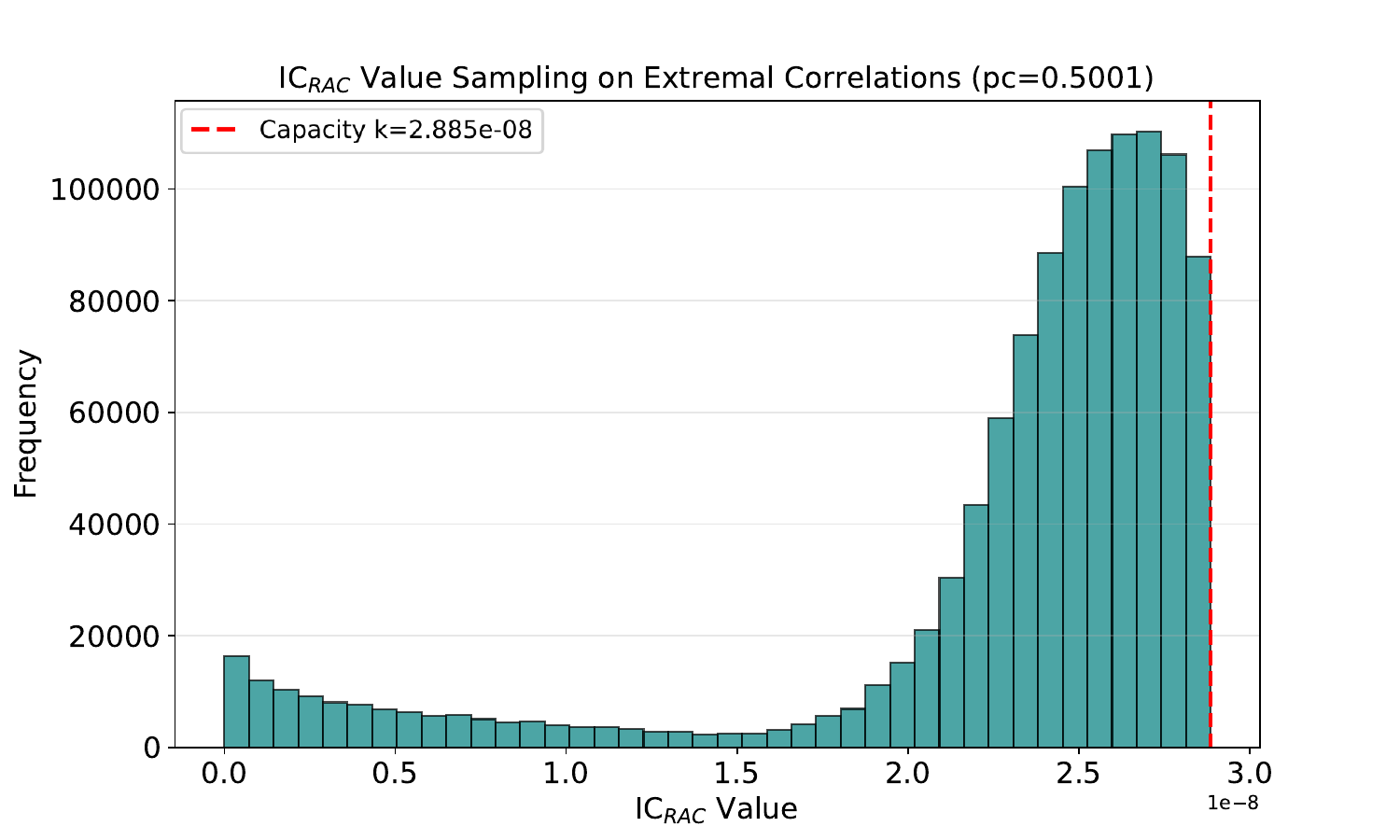}
     \caption{Distribution of $IC_{\mathrm{RAC}}$ values.}
        \label{sub:rac}
    \end{subfigure}
    
    \caption{Distributions of (a) $IC_{\mathrm{red}}$ and (b) $IC_{\mathrm{RAC}}$ values over the sampled extremal correlations. 
    The  $IC_{\mathrm{red}}$ values are bounded by the channel capacity $k$, with a lower cutoff at $1.4 \times 10^{-8}$. In contrast, $IC_{\mathrm{RAC}}$ values show a higher concentration near $k$ but extend down to zero.}
    \label{Ext}
\end{figure}

More specifically, any 2-2-2 extremal quantum correlation can be produced by measuring two-qubit state $\ket{\phi_\theta}=\cos\theta\ket{00}+\sin\theta\ket{11}$
with observables $A_i=\cos a_i\,\sigma_z+\sin a_i\,\sigma_x,\quad
B_j=\cos b_j\,\sigma_z+\sin b_j\,\sigma_x$ ($i,j\in \{0,1\}$) which satisfy fully alternating condition \cite{barizien2025quantum}. 
So in the first step, 
we generated a set of extremal correlation points by picking parameters $\theta\in[0,\pi/4]$, $a_0, b_0, a_1, b_1\in[0,\pi)$ in their respective intervals 
with a step $N=40$, and checking the fully alternating condition. 
We calculated the $IC_{\textrm{red}}$ values for sampled correlation points, and showed their distribution in Fig.~\ref{Ext}(a). 
As seen in the figure, 
the majority of $IC_{\textrm{red}}$ values are distributed near the boundary of channel capacity $k$, while no violation is found. This indicates that $IC_{\textrm{red}}$ can be effective on bounding 2-2-2 quantum correlations. 
For comparison, $IC_{\textrm{RAC}}$ is computed over the same set of extremal correlations and presented in Fig.~\ref{Ext}(b). 
While $IC_{\textrm{RAC}}$ exhibits a more concentrated distribution near $k$, $IC_{\textrm{red}}$ captures the potential information more properly in a part of cases (the $IC_{\textrm{red}}$ values remain above $1.4\times 10^{-8}$ while plenty $IC_{\textrm{RAC}}$ values distribute near $0$).

\begin{figure}[h]
    \centering
  \includegraphics[width=0.49\linewidth]{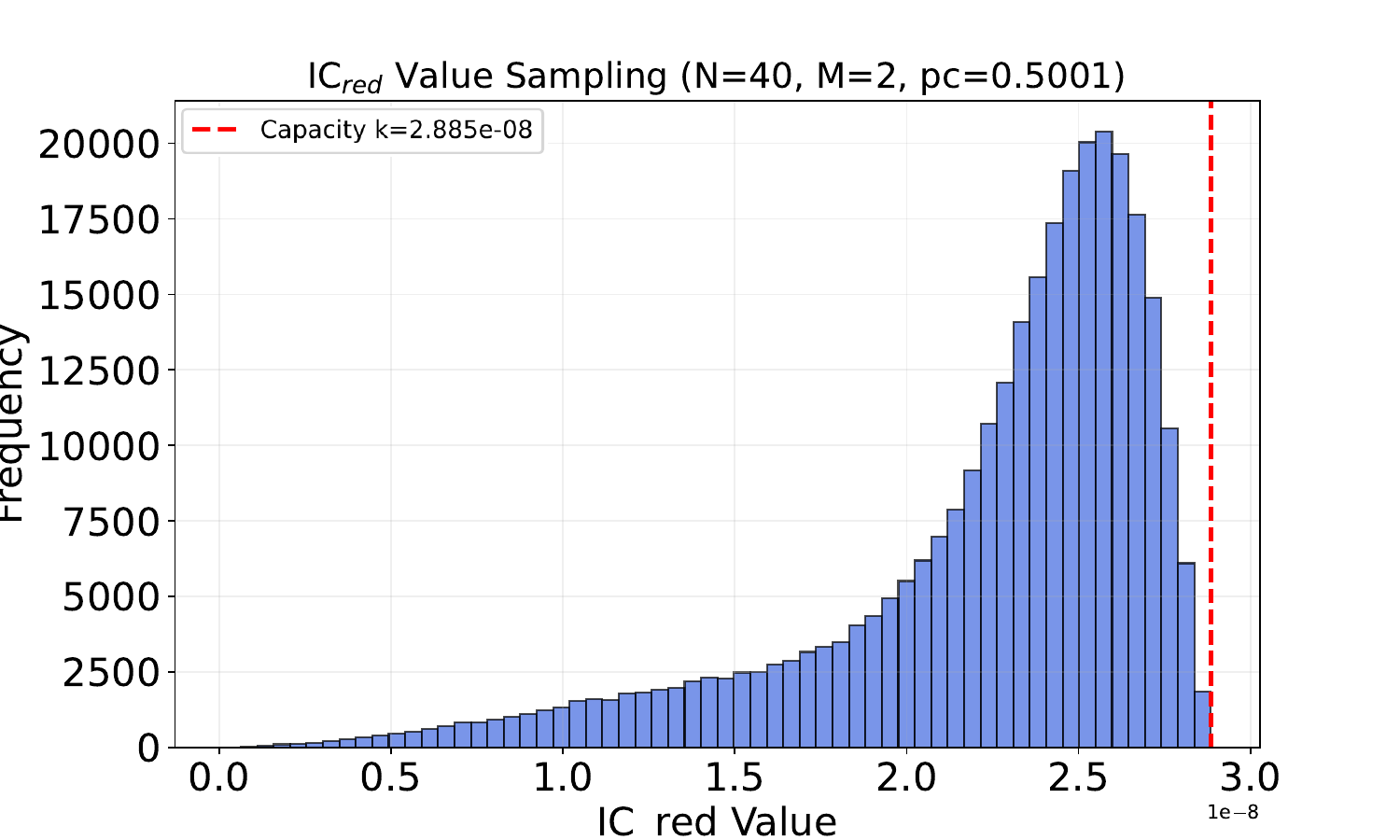}
  \includegraphics[width=0.49\linewidth]{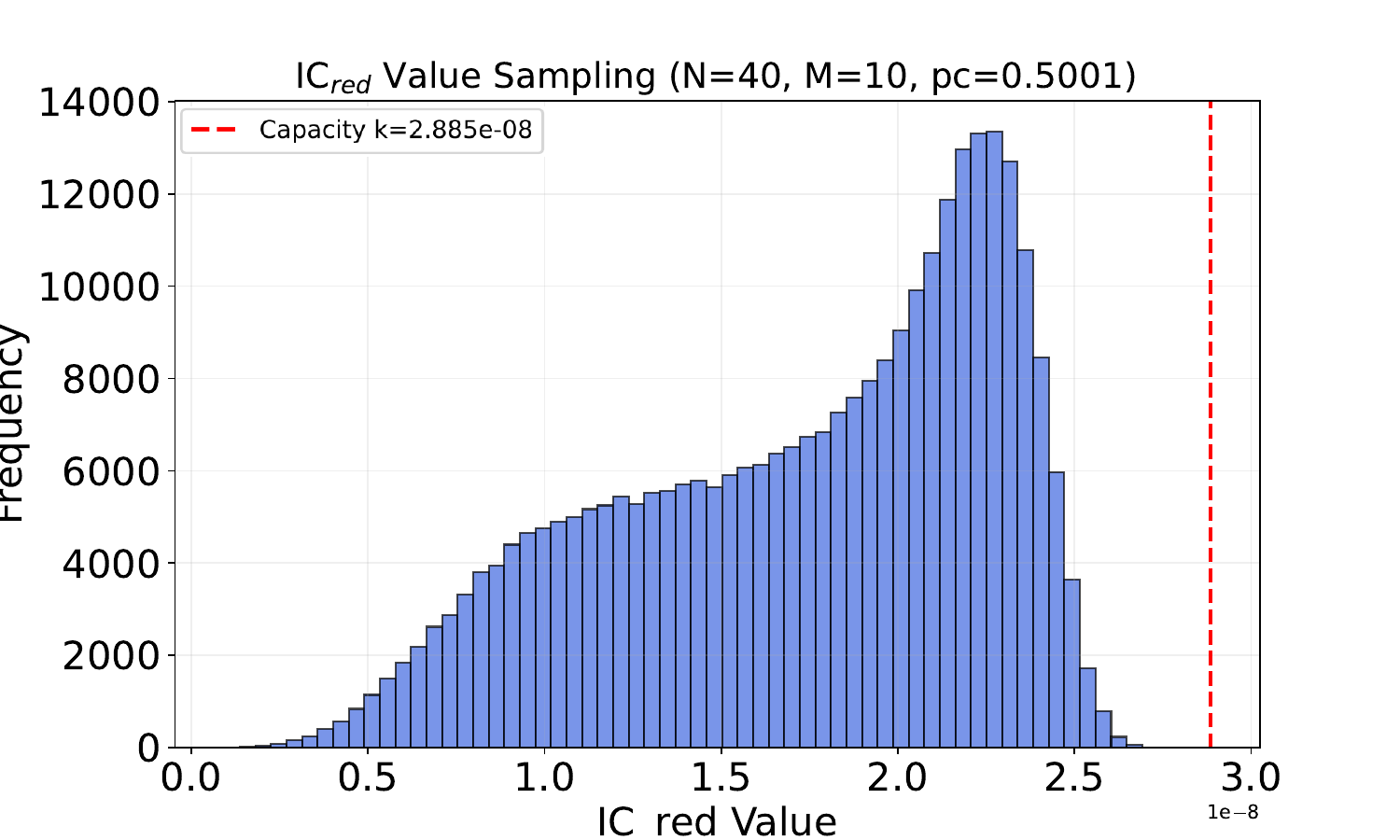}
 \includegraphics[width=0.49\linewidth]{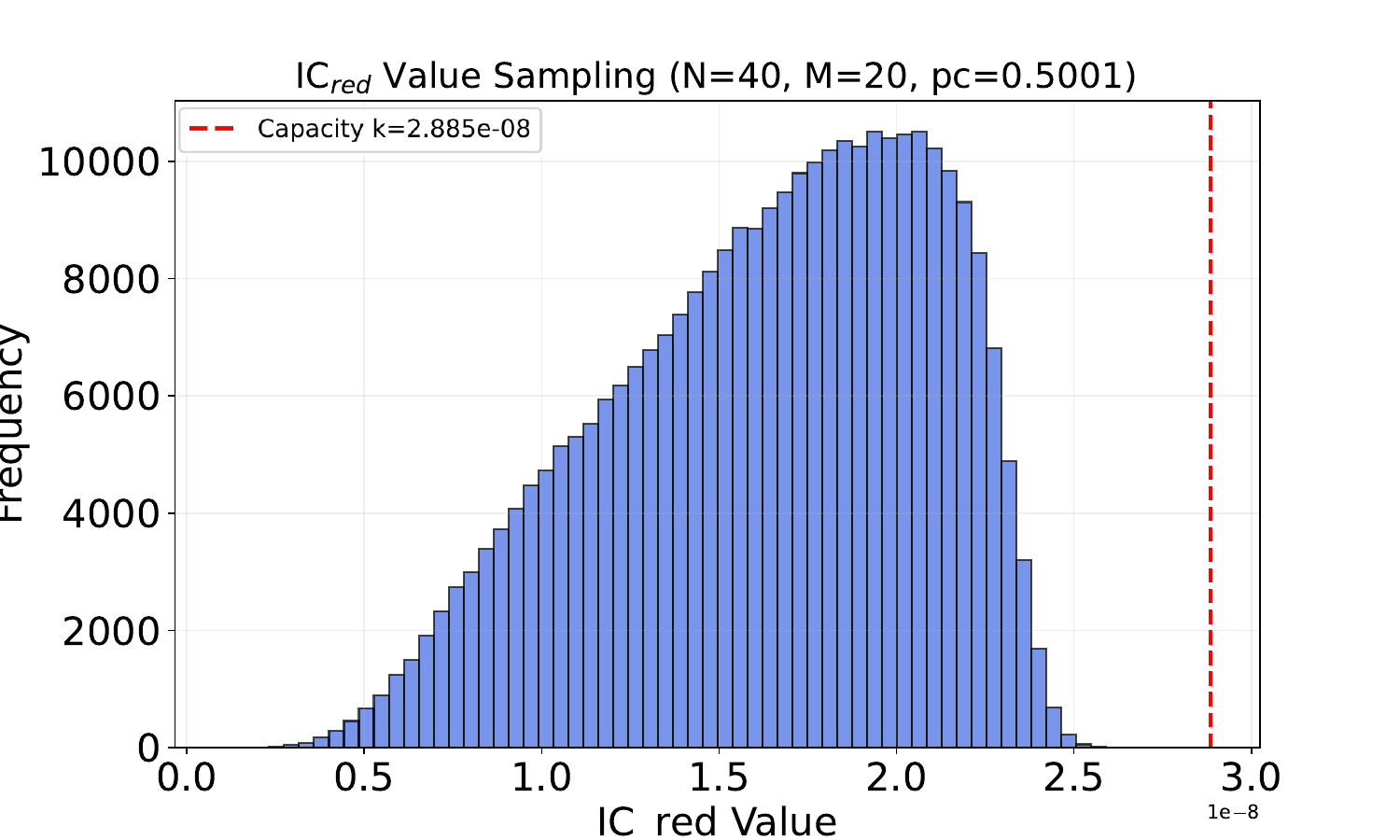}
  \includegraphics[width=0.49\linewidth]{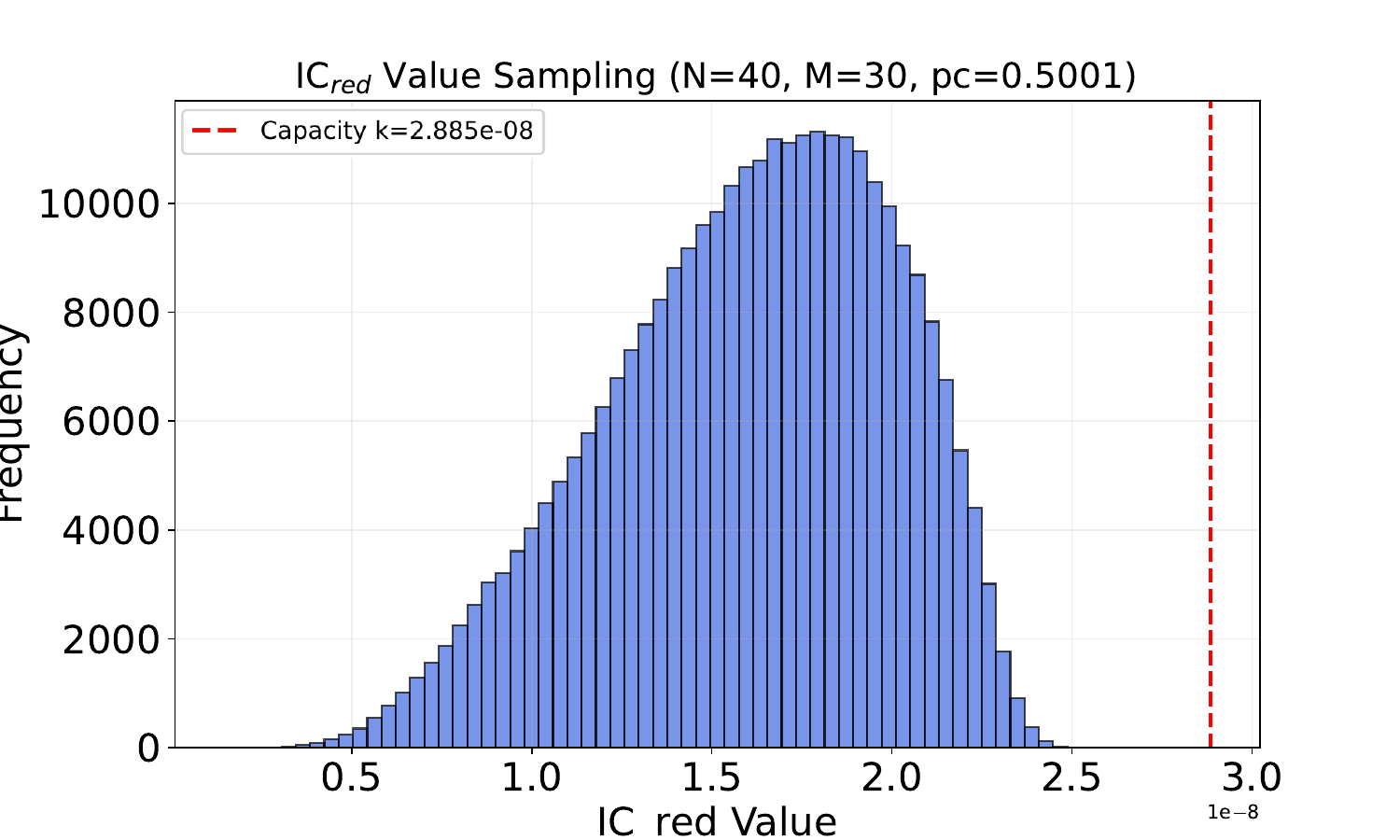}
\includegraphics[width=0.49\linewidth]{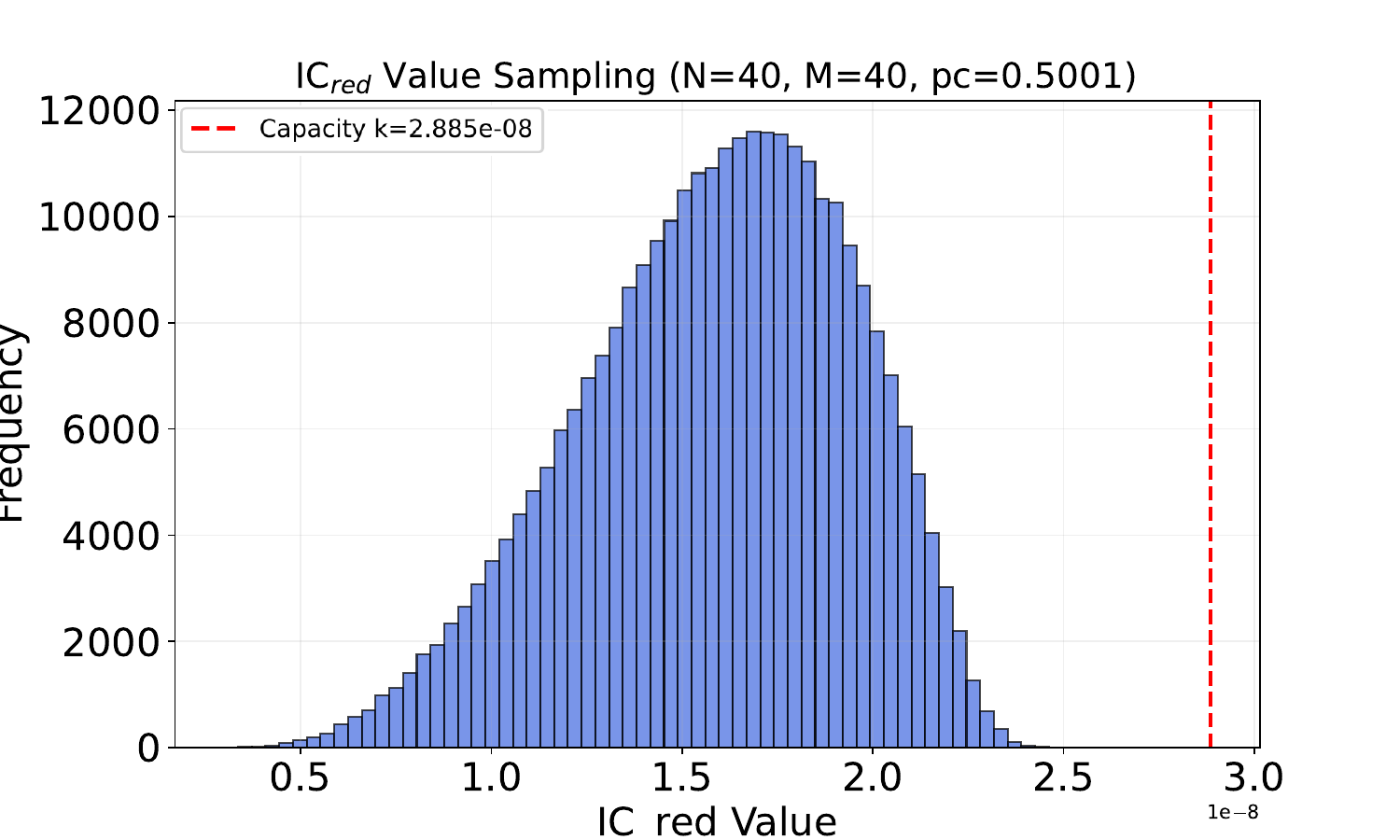}
\includegraphics[width=0.49\linewidth]{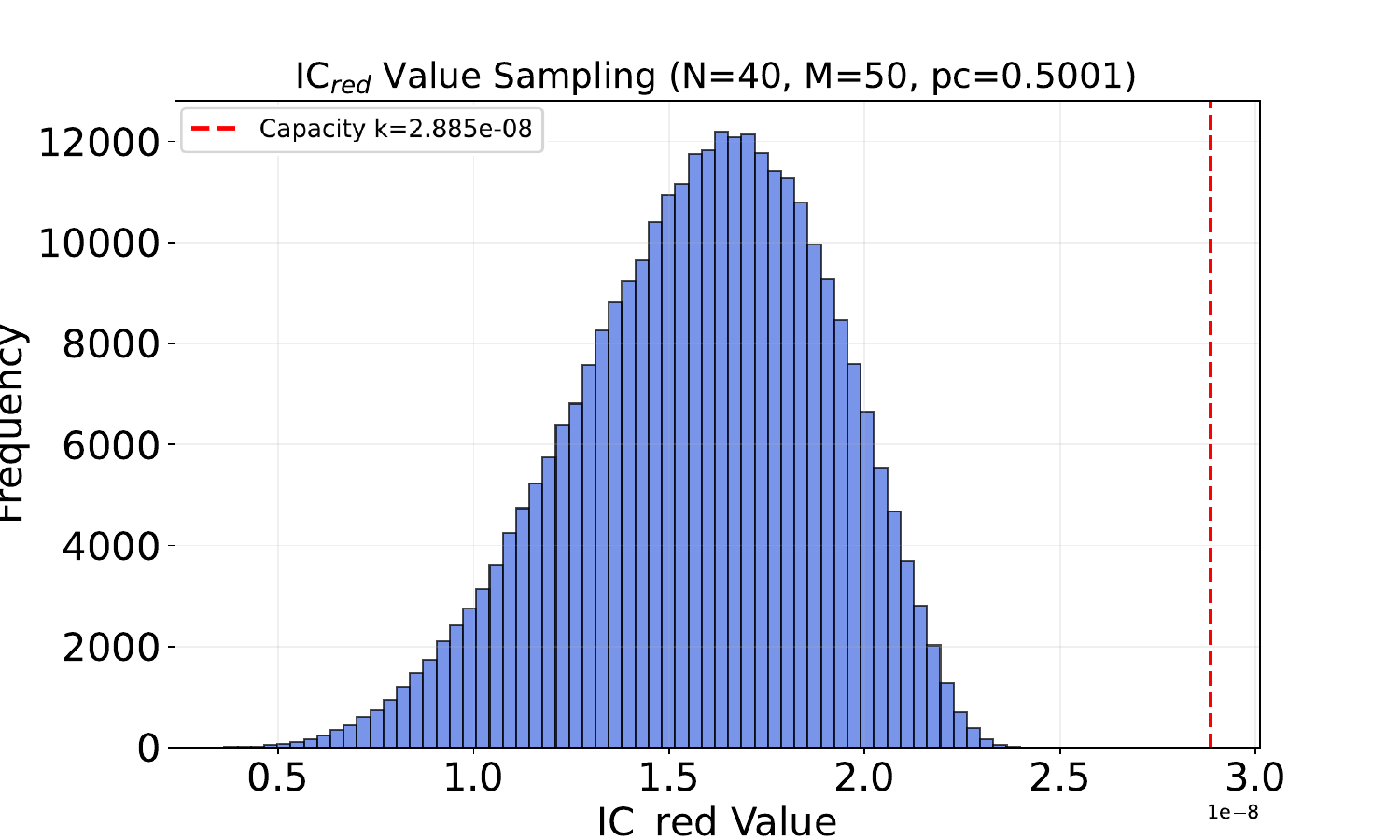}

 \caption{Frequencies of $IC_{red}$ values 300000 random mixture for $M=2,10,20,30,40,50$ random extremal points. We see that when the number of states to be mixed increases, the IC values tend to become smaller as a collective behavior. And when $M\geq 40$, the distribution becomes stable.}
  \label{Extmix}
  \end{figure}

In the second step, 
we made a random sampling of $M$ points in the set of extremal points generated in the first step, and produced correlations by random mixture the $M$ sampled points. 
We tested the cases where 
$M=2,10,20,30,40,50$ respectively,
and in each case we made 300000 random samples. 
We calculated the $IC_{red}$ values corresponding to each correlation point. The results are shown in Fig.~\ref{Extmix}. 
From the figures we see that when M increases,
the distribution of $IC_{red}$ values moves collectively from $k$ toward zero, and when $M\geq 40$, the distributions become stable in spite of the increase of $M$. 
In all cases, no violation of IC occurs. In the appendix \ref{apdxA}, we give more details about the sampling.

In our second approach, we performed a heuristic optimisation of the $IC_{\textrm{red}}$ value over the sub-manifold of physical parameters $(\theta, a_{0}, b_{0}, a_{1}, b_{1})$ constrained by the fully alternating condition. This allows us to identify the maximal $IC_{\textrm{red}}$ value (whether a global or local optimum) attainable by extremal 2-2-2 quantum correlations. This value is then compared directly against the channel capacity $k$. The protocol to map correlations to $IC$ values remains unchanged from our previous description.

The optimisation framework is also applicable for mixtures of $M$ extremal correlations, where the mixing weights are included as optimization variables.
Using this approach, we optimised the $IC_{\textrm{red}}$ values for $M \in \{1, 2, 3, 4, 5\}$. 
To characterise the resulting mixedness, we define the purity $\gamma$ as:
\begin{equation}
    \gamma := \sum_{i=1}^{M} w_{i}^{2},   
\end{equation}
where $\{w_{i}\}$ are the mixing weights corresponding to the optimised $IC_{\textrm{red}}$. 
The metric $\gamma$ indicates whether the maximum is attained by an extremal correlation ($\gamma \approx 1$) or a mixed one ($\gamma < 1$). 

For $M=1,2,3,4,5$, the heuristic optimisation yields $IC_{\textrm{red}}=2.88539\times 10^{-8}$, numerically equal at that precision to the value of $k$. The same optimisation also indicates that this value is obtained for $\gamma=1$, i.e.~on extremal points.

From the numerical results in two approaches, we obtained reliable evidence that Eq.~\eqref{modifiedIC} is obeyed by all 2-2-2 quantum correlations. 
The result also suggest a strong possibility that the $IC_{\textrm{red}}$ expression is quasi-convex. 
Altogether, these results strongly support the reliability of the improvement on the correlation boundaries (shown in Fig.~\ref{fig:results}).

\textit{Conclusion.--} In this paper we proposed a new quantifier for IC principle based on the notion of ``redundant information", and provided an explicit proof of its advantage in the simplest Bell scenario.

In IC, Alice sends information about her random variable $A$ to Bob through a classical channel, and Bob has multiple indicators that can extract information from it. Previous works had captured the principle in the context of a random access code, with $A$ a string of independent symbols, and the desideratum that each of these symbols should be in one-to-one correspondence with Bob's indicators. Our main contribution is to liberate IC from this constraining structure. We highlight that all that IC needs is a method to integrate the pieces of information obtained by different indicators of Bob, while making sure that no information is calculated repeatedly.

A crucial and challenging task to do about the redundant IC formalism  is finding out 
a rigorous and protocol-independent proof, either analytical or numerical, to confirm that all quantum correlations obey such IC principle \footnote{The proof proposed in a previous version of this work was incorrect because it used a definition of the Markov condition that does not hold in the presence of quantum information.}. It would also be interesting to analytically check if our IC expression is quasi-convex. Some more advanced tools such as information geometry might be required to study these problems.

Another bottleneck for further progress comes from classical information theory: there is no suitable candidate expression for redundant information beyond the case of two indicators for Bob, the one we used for our explicit examples. We also notice that, even with our improved approach, there remains a gap between the set of quantum correlations and the set of correlations that violate IC. We conjecture that such gaps can be closed, or at least reduced, by future improvements on the definition of redundant information, or more generally by tightening the quantifier of ``potential information''.

\section*{Acknowledgments}

We thank Francesco Buscemi, Marcin Paw{\l}owski and Dong Yang for feedback and comments, and Koon Tong Goh for sharing his numerical results on the almost-quantum set.

B.Y.~acknowledges the support of the GuangDong Basic and Applied Basic Research Foundation (Grant No. 2024A1515012405). V.S.~is supported by the National
Research Foundation, Singapore through the National
Quantum Office, hosted in A*STAR, under its Centre for
Quantum Technologies Funding Initiative (S24Q2d0009).

\bibliography{ref}

\appendix

\section{Sampling and Heristic Optimisation of IC Values}\label{apdxA}

\subsection{Sampling}
In this appendix we give more details about our numerical approach 1, i.e., the sampling of $IC_{\textrm{red}}$ values over 2-2-2 quantum correlations. 

We begin by introducing the parametrisation of extremal 2-2-2 quantum correlations, 
which was proposed in Ref.~\cite{barizien2025quantum}.
Consider a pure bipartite entangled state $|\phi_{\theta}\rangle = \cos\theta|00\rangle + \sin\theta|11\rangle$, where $\theta \in [0, \pi/4]$. Let $A_x$ and $B_y$ $(x,y\in\{0,1\})$ denote Alice and Bob's binary-outcome measurement operators. The operators $A_x$, $B_y$ can be represented by angle parameters $a_{x}$, $b_{y}$ as:
\begin{equation}
\begin{split}
 &A_{x} = \cos a_{x} \sigma_{z} + \sin a_{x} \sigma_{x}, \\
   &B_{y} = \cos b_{y} \sigma_{z} + \sin b_{y} \sigma_{x}.      
\end{split}
\end{equation}
By measuring $|\phi_{\theta}\rangle$ with all $A_0$, $B_0$, $A_1$, $B_1$, a specific correlation point is generated. That is, each parameter setting $(\theta, a_0, a_1, b_0, b_1)$ characterises a correlation point.

For a given measuremeng angle $a_{x}$ in the $(\sigma_z, \sigma_x)$ plane, 
a steered angle $\tilde{a}_x^s$ is can be defined by a steering transformation as:
\begin{equation}
    \tilde{a}_{x}^{s} = 2 \arctan\left(\tan(a_x/2) \tan^s(\theta)\right)
\end{equation}
where $s \in \{\pm 1\}$.  These steered angles characterize the effective directions of Alice's measurements when viewed through the shared entangled state $|\phi_{\theta}\rangle$.

The boundary of 2-2-2 quantum correlations is fully characterized by $(\theta, a_0, a_1, b_0, b_1)$ which satisfies the \textit{fully alternating condition} \cite{barizien2025quantum}: 
\begin{equation}\label{fullyalternating}
    0 \leq [\tilde{a}_0^s]_\pi \leq b_0 \leq [\tilde{a}_1^t]_\pi \leq b_1 < \pi,
\end{equation}
where $[\alpha]_\pi \equiv \alpha \pmod{\pi}$ and $\tilde{a}_x^s$ are the steered angles of $a_0$ and $a_1$.

Therefore, in step 1, we discretise the physical parameter space $(\theta, a_0, a_1, b_0, b_1)$ using a high-resolution grid with step $N = 40$ for each parameter. A library of extremal points $\{P_{ext}\}$ is then constructed by filtering all discretised parameters with fully alternating condition. The number of extremal correlation points  generated in the library is 1114534.

In step 2,
to explore the interior of the quantum set, we generate $300000$ samples of mixed correlations $P_{mixed}$ by making random convex combination of $M\in{2,5,10,20,30,40,50}$ randomly selected extremal vertices from our library respectively:
\begin{equation}
    P_{mixed} = \sum_{i=1}^M w_i P_{ext}^{(i)}.
\end{equation}
To ensure an unbiased sampling of the convex hull, the mixing weights $\mathbf{w} = (w_1, \dots, w_M)$ are sampled from a Dirichlet distribution with concentration parameters $\bm{\alpha} = (1, \dots, 1)$. This choice generates a uniform distribution over the $(M-1)$-simplex formed by the chosen vertices, filling more and more volume of the quantum set as $M$ increases.

For each sampled correlation, the $IC_{\textrm{red}}$ value is computed for a noisy channel with error probability $p_c = 0.5001$ based on the chosen protocol.
Probabilities are clipped at $10^{-12}$ to prevent logarithmic singularities. This effectively sets our numerical noise floor at $10^{-12}$, well below the targeted channel capacity $k \approx 2.885 \times 10^{-8}$.

\subsection{Optimisation}

We employ the Differential Evolution (DE) algorithm to navigate the non-convex $IC_{\textrm{red}}$ functional. The search space consists of $5M + M - 1$ variables for mixtures of $M$ correlations, where the $5M$ variables correspond to the state and measurements, the $M-1$ variables correspond to the mixing weights.
The optimization is constrained by a penalty function $V$ which measures the deviation from the fully alternating condition:
\begin{equation}
\begin{split}
   V = \min_{s,t \in \{\pm 1\}} [ \text{pos}(-[\tilde{a}_{0}^{s}]_{\pi})+ \text{pos}([\tilde{a}_{0}^{s}]_{\pi} - b_0)\\
   + \text{pos}(b_0 - [\tilde{a}_{1}^{t}]_{\pi}) + \text{pos}([\tilde{a}_{1}^{t}]_{\pi} - b_1) ], 
\end{split}
\end{equation}
where $\text{pos}(x) = \max(0, x)$. The convergence criteria of our optimisation is chosen to be $10^{-15}$, which is close to the limits of double-precision arithmetic.

\end{document}